\documentclass[aip,
               reprint,
               amsmath,
               amssymb,
               ]{revtex4-2}

\usepackage{graphicx}
\usepackage{dcolumn}
\usepackage{bm,bbold}

\usepackage{dsfont}
\usepackage{physics}

\usepackage{xcolor}
\usepackage{hyperref}
\hypersetup{
  colorlinks=true,
    citecolor=violet,
    filecolor=violet,
    linkcolor=violet,
    urlcolor=violet,
    bookmarksopen=true,
    linktoc=page
 }
\usepackage[capitalise]{cleveref}

\usepackage[
	font={sf,small},
	labelfont=bf,
   ]{caption}
\usepackage[
	font={sf,small},
	labelfont=bf,
   ]{subcaption}

\renewcommand{\k}{\kappa}
\newcommand{\x}{\bm{x}}
\renewcommand{\u}{\bm{u}}
\renewcommand{\v}{\bm{v}}
\renewcommand{\exp}[1]{\mathrm{e}^{#1}}
\newcommand{\Steff}{St_{\mathrm{eff}}}
\newcommand{\DNSA}{\textsf{DNS1}}
\newcommand{\DNSB}{\textsf{DNS5}}
\newcommand{\DNSC}{\textsf{DNS0.2}}
\newcommand{\LESA}{\textsf{FDNS1}}
\newcommand{\LESB}{\textsf{FDNS5}}

\newcommand{\etal}{\emph{et al.}}
\newcommand{\Z}{\mathbb{Z}}
\newcommand{\A}{\mathbb{A}}
\renewcommand{\tr}{\mathrm{tr}}
\renewcommand{\Re}{Re}

\begin{document}

\title{
	Dissecting inertial clustering and sling dynamics in high-Reynolds number particle-laden turbulence
}

\author{Lukas A.
	Codispoti} \email{lukasco@ethz.ch} \author{Daniel W.
	Meyer}
\author{Patrick Jenny}
\affiliation{
	Institute of Fluid Dynamics, ETH Z\"urich, Sonneggstrasse 3, 8092 Z\"urich, Switzerland
}
\date{\today}

\begin{abstract}
	In this work, we aim to deepen the understanding of inertial clustering and the
	role of sling events in high-Reynolds number ($\Re$) particle-laden turbulence.
	To this end, we perform one-way coupled particle tracking in flow fields
	obtained from direct numerical simulations (DNS) of forced homogeneous
	isotropic turbulence.
	Additionally, we examine the impact of filtering utilized in large eddy
	simulations (LES) by applying a sharp spectral filter to the DNS fields.
	Our analysis reveals that while instantaneous clustering through the centrifuge
	mechanism explains clustering at early times, the path history effect--the
	sampling of fluid flow along particle trajectories--becomes important later on.
	The filtered fields expose small-scale fractal clustering that cannot be
	predicted by the instantaneous flow field.
	We show that there exists a filter-effective Stokes number that governs the
	degree of fractal clustering and preferential sampling, revealing
	scale-similarity in the spatial distributions and fractal dimensions.
	Sling events are prevalent throughout our simulations and impose prominent
	patterns on the particle fields.
	In pursuit of investigating the sling dynamics, we compute the relative
	velocity, ensemble-averaged over proximal neighboring particles, to identify
	particles undergoing caustics.
	As postulated in recent theories, we find that in fully resolved, high-$\Re$
	turbulence, sling events occur in thin sheets of high strain, situated between
	turbulent vortices.
	This behavior is driven by rare, extreme events of compressive straining,
	manifested by fluctuations of the flow velocity gradients that propagate back
	and forth the positive branch of the Vieillefosse line.
\end{abstract}

\maketitle

\section{Introduction}\label{sec:intro}
Inertial particles detach from the flow and form clusters.
This phenomenon is significant in a wide range of natural processes and
engineering applications, e.g., the growth of droplets in atmospheric
clouds~\cite{grabowskiGrowthCloudDroplets2013}.
Even in randomly-forced suspensions with a single characteristic timescale,
intricate patterns of the density field can be
observed~\cite{becFractalClusteringInertial2003a}.
At high Reynolds numbers, turbulent flows exhibit broad ranges of highly
separate temporal and spatial scales with strong intermittency, especially at
high wavenumbers.
Clustering is then a complex, multiscale process, driven by multiple
mechanisms.
Particles cluster both by action of the instantaneous flow field and by the
path-history effect, i.e., by collecting memory along their paths.
The former is typically described as particles being centrifuged out of regions
of high vorticity and accumulating in regions of high
strain~\cite{maxeyGravitationalSettlingAerosol1987,squiresPreferentialConcentrationParticles1991}.
The latter produces fractal spatial distributions and leads to sling events,
whereby the particle dynamics exhibits phase-space singularities, so-called
caustics~\cite{falkovichAccelerationRainInitiation2002,wilkinsonCausticsTurbulentAerosols2005,gustavssonStatisticalModelsSpatial2016},
resulting in anomalous particle number density fields and large relative
velocities.
Caustics are said to play a key role in the collisions of droplets in
clouds~\cite{wilkinsonCausticActivationRain2006,pumirCollisionalAggregationDue2016},
which initiate rainfall, offering a possible solution to the bottleneck problem
of cloud physics\cite{deepuCausticsinducedCoalescenceSmall2017}.

It is widely accepted (e.g.,
Refs.~\onlinecite{cenciniDynamicsStatisticsHeavy2006,becHeavyParticleConcentration2007,rayPreferentialConcentrationRelative2011},
to only name a few) that clustering is most intense when flow and particle
timescales are comparable, i.e. when the Stokes number, defined as the ratio
between particle relaxation to the Kolmogorov timescale, $St=\tau_p/\tau_\eta$,
is close to unity.
This implies that the smallest scales of the flow are most relevant in
explaining clustering.
In this work, we investigate the effect of filtering the turbulent flow field,
i.e., removing the smallest scales, on the mechanisms that drive particle
clustering.
This is highly relevant for LES of particle-laden turbulence, in which the
Navier-Stokes equations are solved for a filtered, coarse-grained velocity
field, leaving the sub-grid scales (SGS) unresolved.
While this approach significantly lowers computational costs, it requires
modeling to capture the influence of the unresolved scales.
This poses a challenge, especially for particle-laden flows, where clustering
is driven by the coherent motion at the smallest scales.

In this work, we assume dense point-particles of negligible spatial extensions
in dilute suspensions,
so that their motion can be approximated by~\cite{gustavssonStatisticalModelsSpatial2016}
\begin{equation}
	\label{eqn:pp} \dot{\x} = \bm{v}, \quad
	\dot{\bm{v}} = \frac{\bm{u}(\bm{x},t) - \bm{v}}{\tau_p},
\end{equation}
where $\bm{x}$ and $\bm{v}$ are the particle position and velocity, respectively,
and $\bm{u}(\bm{x},t)$ is the local flow velocity.
\Cref{eqn:pp} neglects particle-fluid and particle-particle interactions,
which is commonly referred to as \emph{one-way coupling}.
The approach is convenient for DNS, since it separates the flow computation
from the particle tracking and enables the study of clustering in isolation.
If there exists a continuous, smooth velocity field that describes
the motion of the particles, $\u_p(\x,t)$, then \cref{eqn:pp} can be written in the Eulerian frame of reference as
\begin{equation}
	\label{eqn:up}
	\frac{\mathrm{D}_{p}\u_p}{\mathrm{D}t} :=
	\pdv{\u_p}{t} + (\u_p\cdot\nabla)\u_p =
	\frac{\bm{u} - \u_p}{\tau_p}.
\end{equation}
Since $\dot{\v}$ in \cref{eqn:pp} represents the substantial derivative of
that particle velocity field along a particle path, i.e., ${\mathrm{D}_{p}\u_p}/{\mathrm{D}t}$,
the particle number density $n(\x,t)$ is a conserved
scalar that is transported with $\u_p$ and satisfies
\begin{equation}\label{eqn:cons}
	\frac{\mathrm{D}_{p} \ln n }{\mathrm{D}t}
	= - \nabla\cdot\u_p := -\xi,
\end{equation}
which can be derived from the conservation equation
$\partial_t n + \nabla\cdot(n\mathbf{u}_p) = 0$.
The particle velocity divergence, herein after denoted by the symbol $\xi$, is
an explicit source term on the right-hand side of \cref{eqn:cons}.
In other words, it is the compressibility of the particles which lag the
\emph{in}compressible flow that produces variance in $n$.
From \cref{eqn:up} we find that $\xi$ evolves according to
\begin{equation} \label{eqn:xi}
	\frac{\mathrm{D}_{p}\xi}{\mathrm{D}t}
	= - \frac{\xi}{\tau_p} - \tr(\mathbb{U}^2),
\end{equation}
where $\mathbb{U}_{ij}=\tfrac{\partial u_{p,i}}{\partial x_j}$ (and $\xi=\tr(\mathbb{U})$).
As seen from \cref{eqn:xi}, the particle velocity divergence relaxes at the
rate $\tau_p^{-1}$ towards zero, since the carrier fluid is incompressible,
i.e., $\nabla\cdot\u=0$.
The non-linear term on the right-hand side of \cref{eqn:xi} can, however,
render the system unstable.
This is a first indication that at large enough Stokes numbers the particle
dynamics are susceptible to finite-time singularities, i.e., that $\xi$ might
escape to $-\infty$ in finite time.
In fact, as $\tau_p\rightarrow\infty$, \cref{eqn:up} reduces to the
three-dimensional (3D) inviscid Burgers equation, which is well-known for
producing shocks resulting from gradient \emph{self-amplification} of the
convective term~\cite{johnsonMultiscaleVelocityGradients2024}.
In the context of particle-laden flows, blow-up events of this kind are
referred to as caustics, which will be discussed in more detail below.

Let us now turn to the mechanisms that drive inertial particles to cluster.
Maxey~\cite{maxeyGravitationalSettlingAerosol1987} used the continuum approach
and expressed for $St \to 0$ the particle velocity field as $\u_p = \u +
	\tau_p\tfrac{D\u}{Dt} + \mathcal{O}(\tau_p^2)$.
The particle velocity divergence then results from this limiting expression as
\begin{equation}\label{eqn:maxey}
	\xi \approx 2\tau_pQ.
\end{equation}
Here, $Q = -\tfrac{1}{2}\tr(\A^2)$ is the second invariant of the flow velocity
gradient (FVG) tensor $\A_{ij}=\tfrac{\partial u_i}{\partial x_j}$, expressing
the difference between rotation and dissipation of the local flow
topology~\cite{meneveauLagrangianDynamicsModels2011,johnsonMultiscaleVelocityGradients2024}.
Therefore, \cref{eqn:maxey} indicates that particles tend to reside in regions
where strain dominates vorticity, consistent with the notion of \emph{Maxey's
	centrifuge}, according to which particles are ejected out of turbulent eddies
and accumulate in areas of low vorticity.
This mechanism, also referred to as \emph{preferential sampling}, explains
clustering purely based on the instantaneous flow field.
Strictly speaking, it is only valid at small Stokes numbers, whereas higher
order terms of the FVGs play a more significant role for heavier particles.
As a particle's inertia increases, memory effects become more pronounced, and
the history of $\A$ sampled along its trajectory exerts greater influence on
its instantaneous velocity.

Aside from Maxey's centrifuge, the clustering of inertial particles in
turbulence is indicative of the dissipative dynamics governed by \cref{eqn:pp},
which evolve towards a fractal phase-space
attractor~\cite{becFractalClusteringInertial2003a,becMultifractalConcentrationsInertial2005,becStatisticalModelsDynamics2024a}.
This means that particles cluster onto fractal sets in physical space in the
smallest scales of the flow, i.e., at lengths $\ell\sim10\eta$.
The fractal dimension, which quantifies the 'space-fillingness' of the particle
distribution, depends on the Stokes number and has significant implications for
processes such as collision, evaporation or condensation of droplets in
turbulence.
Fractal patterns have been observed in particle-laden simulations with
synthetic as well as fully-resolved turbulence and are commonly quantified with
the correlation dimension\cite{becStatisticalModelsDynamics2024a}, $d_2$.
If $N(r)$ is the number of neighbors a particle is surrounded by inside the
sphere of radius $r$, then~\cite{gustavssonStatisticalModelsSpatial2016}
$N(r)\sim r^{d_2}$.
In 3D configuration space, fractal particle clusters are characterized by
$d_2<3$.
While preferential sampling is still relevant in the small
scales\cite{becHeavyParticleConcentration2007}, the history of the FVGs that
particles sample along their paths must be taken into account in order to
explain fractal clustering.
Here, it is appropriate to move away
from \cref{eqn:xi}, which describes the evolution of the divergence of the
\emph{synthetic}, continuous particle velocity field, and instead study the
evolution of the Lagrangian particle
velocity gradient tensor, $\Z_{ij}=\tfrac{\partial v_{i}}{\partial x_j}$, given by~\cite{falkovichAccelerationRainInitiation2002,gustavssonStatisticalModelsSpatial2016,becStatisticalModelsDynamics2024a}
\begin{equation} \label{eqn:Z}
	\dot{\Z} =  \frac{\A - \Z}{\tau_p} - \Z^2.
\end{equation}
Note that \cref{eqn:xi} corresponds to the trace of \cref{eqn:Z} in the
Eulerian frame of reference.
By studying \cref{eqn:Z}, however, we depart form the restricting assumption of
a continuous particle velocity field and directly observe the effect of the
FVGs, which vanishes when applying the trace operator due to incompressibility.
The path-history effect gives rise to \emph{multiplicative
	amplification}~\cite{gustavssonStatisticalModelsSpatial2016}, referring to the
mechanism by which the volumes spanned by small particle clouds randomly
contract and expand.
The rate of contraction/expansion as well as the spatial Lyapunov exponents of
the system can be expressed using $\tr(\Z)$.

The mechanism that has received arguably the most attention
lately\cite{meibohmPathsCausticFormation2021,meibohmCausticsTurbulentAerosols2023,meibohmCausticFormationNonGaussian2024,bartaCausticsTurbulentAerosols2022,bhatnagarRateFormationCaustics2022,leeIdentificationParticleCollision2023}
is clustering by caustics, which are singularities in the dynamics of $\Z$.
As inertial particles experience individual histories of flow velocity
(gradients) along their trajectories, which influence their instantaneous
state, multiple particles with different velocities can occupy the same
positions in time and space.
This phenomenon was termed \emph{sling effect} by
Falkovich~\etal~\cite{falkovichAccelerationRainInitiation2002}, and the
phase-space singularity required for it to occur became known as a
\emph{caustic}~\cite{wilkinsonCausticsTurbulentAerosols2005}, named after the
patterns that light casts onto the bottom of a body of water on a sunny day.
A caustic is a blow up of \cref{eqn:Z}, corresponding to
$\tr(\Z)\rightarrow-\infty$.
When it occurs, the phase space folds, and the singularity in the particle
velocity divergence leads to multivalued particle velocities and extreme local
increases in the particle number density.

\begin{figure*}
	\includegraphics[width=0.9\textwidth]{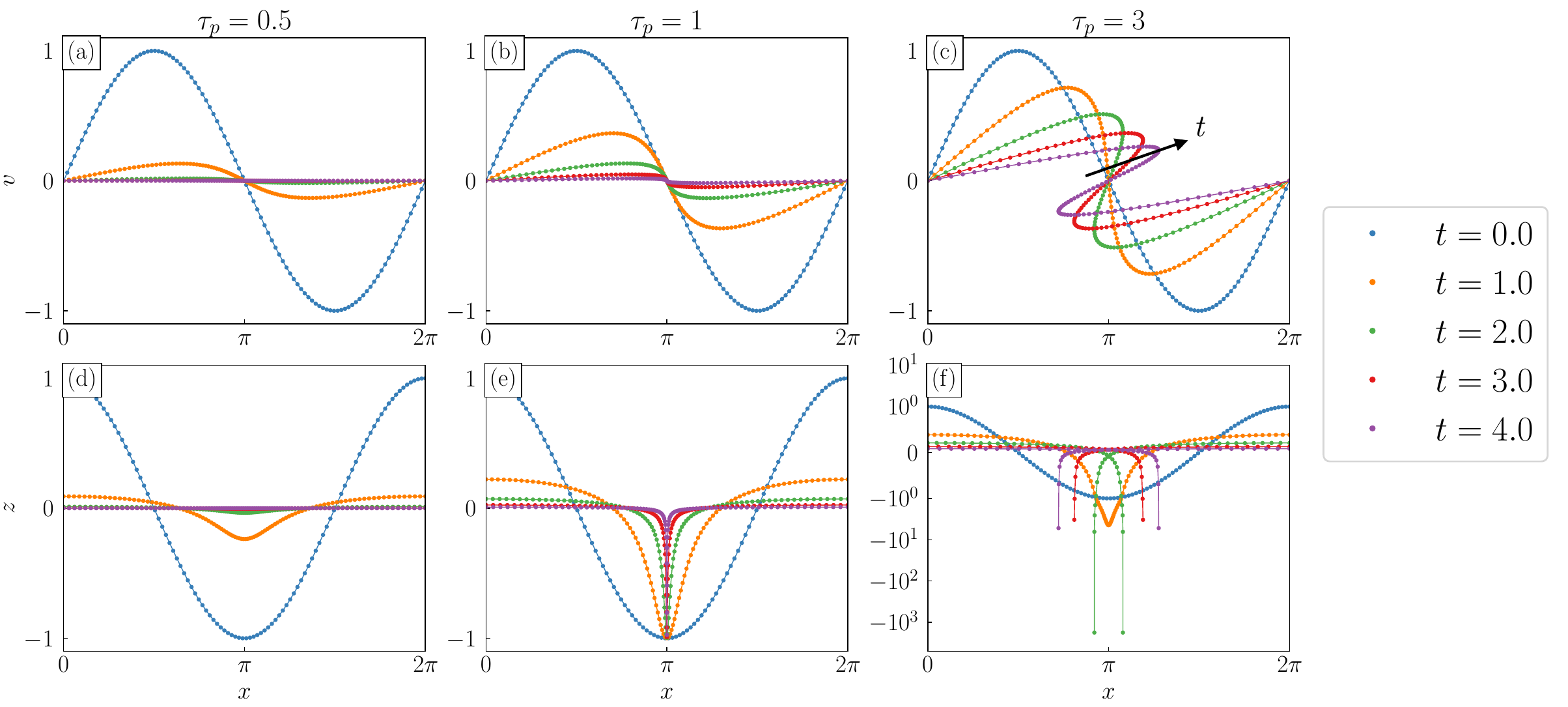}
	\caption{\label{fig:1dcaustics}
		Caustic formation in a one-dimensional model problem.
		Solution to (a-c) $\cref{eqn:1d}$ and (d-f) $\cref{eqn:1dz}$ for 100 Lagrangian
		particles with initial conditions $v=\sin(x)$ and $z=\cos(x)$ (shown in blue).
		To aid visual comprehension, the particles of the same instance in time are
		connected by lines.
	}
\end{figure*}
To illustrate the physical phenomenon of the caustic, let us consider a simple,
one-dimensional setting.
Here, we integrate the equations for the particle velocity
\begin{equation}
	\label{eqn:1d}
	\dot{v} = -\frac{v}{\tau_p}
\end{equation}
(where the shorthand notation $\dot{w}$ for an arbitrary function $w(x,t)$ stands for $\dot{w} = \partial_t w + v \partial_x w$)
and the gradient $z = \pdv{v}{x}$
\begin{equation}
	\label{eqn:1dz}
	\dot{z} =-\frac{z}{\tau_p} - z^2,
\end{equation}
(derived from \cref{eqn:1d} and similar to \cref{eqn:Z}) using 100 point-particles with the initial condition $v=\sin(x)$ and $z=\cos(x)$, respectively.
The motion of the particles is reduced to one dimension and (in contrast to
\cref{eqn:pp}) a constant, zero-valued flow $u=0$ is prescribed.
The results for three different values of $\tau_p$ are displayed in
\cref{fig:1dcaustics}, with the top- and bottom-row panels showing the time
series of solutions to \cref{eqn:1d,eqn:1dz}, respectively.
We observe that light particles (Figs.~\ref{fig:1dcaustics}(a,d))
readily evolve in time towards $v=0$ and $z=0$, respectively.
The solution remains smooth and well-behaved at long integration times and both
signals essentially vanish.
In the intermediate case ($\tau_p=1$), the velocity gradient develops a sharp
peak into the negative direction at $x=\pi$, yet remains bounded
(\cref{fig:1dcaustics}(e).
For $\tau_p=3$, \cref{eqn:1dz} is unstable and $z$ blows up to large negative
values (\cref{fig:1dcaustics}(f)).
A fold in $x-v$ space develops, the hallmark of the caustic singularity.
This is caused by the de-regularizing effect that inertia exerts on the
particle dynamics.
Clearly, the relaxation term on the right-hand side of \cref{eqn:1d} has a
dampening effect on the system: it dictates how fast the solution converges to
$v=0$.
At large $\tau_p$, the relaxation is slow and particles travel far enough to
overtake one another in due time; a fold in phase space ensues.
Caustic singularities can only occur in Lagrangian flows.
The Eulerian equation equivalent to \cref{eqn:1d}, $\pdv{u_p}{t} +
	\pdv{(u_p^2/2)}{x} =-\frac{u_p}{\tau_p}$, on the other hand, corresponds to the
inviscid Burgers equation with a source term on the right-hand side and thus
can lead to shocks, if not sufficiently suppressed by the relaxation term (or
by viscosity), but the weak solution always remains single-valued, whereas the
particles in sling events effectively exhibit multivalued local velocities.

The results shown in \cref{fig:1dcaustics} depict the solution to a distinctly
simple particle-laden flow.
In real turbulence and at high $\Re$, particles are subjected to highly
intermittent, 3D flows, which fluctuate on a range of time and length scales.
\Cref{fig:1dcaustics} illustrates, however, that even in the absence of a
turbulent (or any, for that matter) signal, the dynamics of particles with large
Stokes numbers are susceptible to the formation of caustics.
In the following, we will briefly review recent advances in the field involving
turbulent or statistical model flows.

Whether a caustic forms depends on the history of FVGs that the particles
experience along their trajectories.
Recently, Meibohm and
co-authors~\cite{meibohmPathsCausticFormation2021,meibohmCausticsTurbulentAerosols2023,meibohmCausticFormationNonGaussian2024}
have investigated paths to caustic formation in the so-called \emph{persistent
	limit}, in which particles have weak inertia ($St\ll1$) and see the flow as
quasi-constant ($Ku\gg1$).
The persistence of the flow is characterized by the Kubo
number~\cite{meibohmHeavyParticlesPersistent2019} $Ku=\tau_\eta/\tau_a$, where
$\tau_a=\eta/u_\eta$ is the advection timescale, i.e., the typical time it
takes for a tracer to travel the distance $\eta$.
It was demonstrated that ahead of participating in a sling event, particles
reside in flow regions where the topology is dominated by strain overwhelming
rotation (i.e., $Q<0$) and by strain-rate self-amplification.
The latter is characterized by large positive values of the third invariant of
the FVG tensor, $R = -\tfrac{1}{3}\tr(\A^3)$, which expresses the difference
between enstrophy production and dissipation
production~\cite{meneveauLagrangianDynamicsModels2011}.
In contrast to the vast majority of particles, which expectedly sample $\langle
	Q\rangle=0$ and $\langle R\rangle=0$ along their paths (with $\langle \dots
	\rangle$ representing Lagrangian time averages), particles that undergo
caustics boast large excursions of the invariants ahead in time.
Optimal fluctuations of $Q$ and $R$ that trigger caustics were derived of the
form\cite{meibohmCausticsTurbulentAerosols2023} $Q(t)\sim -(A_{th}f(t))^2$ and
$R(t)\sim (A_{th}f(t))^3$, where $A_{th}$ is an $St$-dependent threshold and
$f(t)$ a strain correlation function, both determined empirically by fitting to
simulation data.

\begin{figure}
	\centering
	\includegraphics[width=\columnwidth]{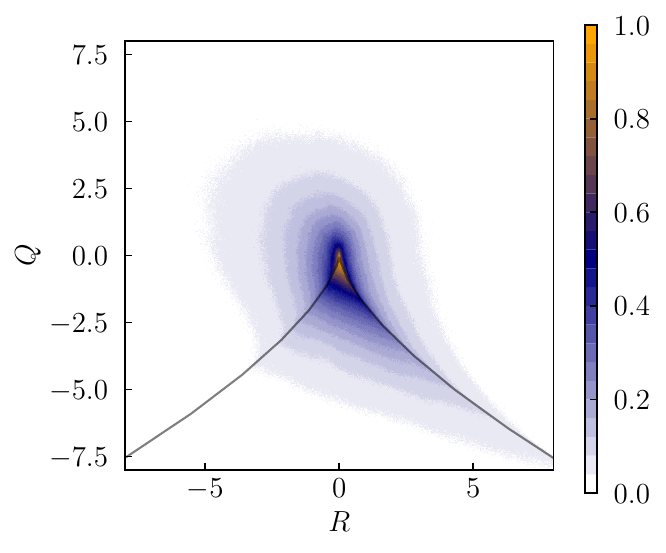}
	\caption{\label{fig:QR-eul}
	Joint probability density of $Q$ and $R$ computed from a filtered DNS field,
	scaled with its maximum value.
	The vertical and horizontal axes are scaled with $A_0$ and $A_0^{-3/2}$,
	respectively, where $A_0$ is an arbitrary constant.
	The solid black line shows the Vieillefosse line.
	Note that the unfiltered flow fields qualitatively produce similar joint
	probability distributions (not shown).
	}
\end{figure}

Here, it is of significance to consider the plane spanned by $Q$ and $R$ in
order to describe the dominant flow topology.
Simulations and experiments of turbulence show that the joint probability
density function (JPDF) $p(Q,R)$ is skewed towards the fourth quadrant of the
$QR$-plane\cite{meneveauLagrangianDynamicsModels2011} ($Q<0$, $R>0$).
There, the strain-rate tensor has two positive (stretching) and one negative
(contracting) eigenvalue, and the flow topology is dominated by strain and
strain production, i.e., strain-rate
self-amplification\cite{johnsonMultiscaleVelocityGradients2024}.
The above-mentioned skewness of the JPDF is evident in \cref{fig:QR-eul}, which
is obtained from the sharp spectrally-filtered flow and also displays the
so-called Vieillefosse line\cite{vieillefosseLocalInteractionVorticity1982}
$(R/2)^2+(Q/3)^3=0$.
Gradients concentrate around its positive branch ($R>0$), along which they
self-amplify and create the most extreme values of
$\A$\cite{johnsonMultiscaleVelocityGradients2024,meneveauLagrangianDynamicsModels2011}.
This exact flow topology that propagates along the positive Vieillefosse tail
was shown by Meibohm~\etal~\cite{meibohmCausticsTurbulentAerosols2023} to drive
particles into sling events in statistical models of particle-laden turbulence.
In real turbulence, on the other hand, caustics are driven by a range of FVGs
that scatter broadly below the Vieillefosse
line~\cite{meibohmCausticFormationNonGaussian2024}, yet still qualitatively
match the predictions from the statistical model.
The results presented in these studies are limited to small Stokes numbers,
specifically $St\lesssim0.3$.
At larger inertia, particles are more susceptible to instabilities, and
non-optimal fluctuations of $\A$ become increasingly likely to cause $\Z$ to
blow up.
It remains unclear to what extent these findings generalize to turbulence at
larger $\Re$, smaller $Ku$ and a wider range of $St$.
Such regimes are relevant, e.g., for the collision of droplets in clouds and
thus deserve attention.

Returning to the Eulerian approach introduced at the beginning of this section,
Lee \& Lee~\cite{leeIdentificationParticleCollision2023} solved \cref{eqn:up}
explicitly to perform DNS of particle density fields advected by 3D isotropic
turbulence at a moderate Reynolds number of $\Re_\lambda=14$.
The simulations were carried out until $\xi$ diverged, which was then
identified as the occurrence of a caustic.
It was observed that sling events take place in thin layers of low vorticity
and high strain, 'packed' in between turbulent vortices.
The flow topologies of these regions coincided with the Vieillefosse line in
the $QR$-plane.

These recent efforts to understand the sling dynamics of inertial particles
underpin the adamant mystery posed by the formation of caustics in turbulent
flows.
The present work aims at further elucidating this phenomenon by providing
observations of sling events of unprecedented detail on the basis of an
extensive DNS campaign of particle-laden turbulence at large $\Re$.
We will verify the aptitude and limits of novel theories of caustics by
identifying and studying sling events in fully-resolved turbulence with
particles of Stokes numbers of up to $St=5$.
To the best of our knowledge, such results have up to now been limited to
statistical flows or turbulence at $Re\leq200$ and $St\leq0.3$.
In addition, we will dissect the clustering mechanisms originating from the
centrifuge and the path-history effect, and we will study the effect of
filtering on the spatial distributions of the particle fields, which has
important implications for modeling particle-laden turbulence in LES.

This paper is structured as follows.
The numerical methods are described in \cref{sec:dns}.
The results are discussed in \cref{sec:results}, divided into three parts:
First, we discuss instantaneous clustering and the relative importance of
Maxey's centrifuge and the history-path effect (\cref{subsec:instant}).
We then turn to the effect of the filter and show that scale-similar patterns
are obtained (\cref{subsec:similarity}).
Thirdly, we cover sling events and investigate the flow topologies that trigger
caustics (\cref{subsec:sling}).
Finally, in \cref{sec:conclusion}, we summarize our findings.

\section{Numerical simulations}\label{sec:dns}
Direct numerical simulations of incompressible homogeneous isotropic turbulence
laden with $N_p = 5.4\times10^9$ mono-dispersed particles are performed.
The turbulent flow fields are obtained from the public database of the Johns
Hopkins University~\cite{liPublicTurbulenceDatabase2008}.
They are produced by a pseudo-spectral code with low-wavenumber forcing in a
cubic domain of length $L=2\pi$ with periodic boundary conditions.
The Fourier and physical space are discretized into $2\k_{\max}=1024$
wavenumbers per dimension and $N_g=1024^3$ grid points, respectively.
The Taylor-scale Reynolds number is $Re_\lambda=418$ and the Kolmogorov length
scale is $\eta=0.0028$.
The simulation parameters are provided in \cref{tab:table1}.
The fields are given in non-dimensional units.
Time is measured in terms of the DNS time, which is provided in intervals of
$\Delta t=0.002$.
The method to calculate $\tau_\eta$ after applying the filter and the
definition of the filter-effective Stokes number $\Steff$ are provided in
\cref{subsec:similarity}.
\begin{table}
	\caption{\label{tab:table1}
		Simulation parameters.
	}
	\begin{ruledtabular}
		\begin{tabular}{c c c c c c c}
			Case &$\k_{c}$&$\tau_\eta$&$\tau_p$&$Ku$&$St$&$\Steff$\\ \DNSC & 512 & 0.0424 &
			0.0085 & 10.3 & 0.2 & - \\ \DNSA & 512 & 0.0424 & 0.0424 & 10.3 & 1 & - \\
			\DNSB & 512 & 0.0424 & 0.2120 & 10.3 & 5 & - \\ \LESA & 16 & 0.2165 & 0.0424 &
			21.4 & - & 0.2 \\ \LESB & 16 & 0.2165 & 0.2120 & 21.4 & - & 0.98\\
		\end{tabular} \end{ruledtabular} \end{table}

We use an
in-house parallel solver to integrate \cref{eqn:pp} in time for each particle
using a simplified version of the scheme by
Jenny~\etal~\cite{jennySolutionAlgorithmFluid2010} The fluid velocity is
evaluated at the particle positions by trilinear interpolation.
Ireland~\etal~\cite{irelandHighlyParallelParticleladen2013} report that in
comparison to spectral interpolation, errors of up to $1\%$ are incurred with
linear interpolation schemes.
We view this error as an upper bound to the present work, since a higher
resolution is used.
Interpolation errors are accepted; it was, however,e verified that the method
is conservative, i.e., that tracers do not cluster.

A sharp spectral filter with a cut-off wavenumber $\k_c$ is used to truncate
the turbulent energy spectrum such that the filtered velocity field is given by
$\bar{\u}(\x,t) = \sum _{\bm{\k}:|\bm{\k}| < \k_c}
	\bm{\hat{u}}(\bm{\k},t)\exp{i\bm{\k}\cdot\x}$, where $\bm{\hat{u}}(\bm{\k},t)$
is the Fourier mode of the flow velocity with wavenumber vector $\bm{\k}$.
The filter effectively removes all turbulent length scales smaller than $\ell_c
	= \pi/\k_c$.
We note that the filtered velocity fields are distinctly different from
synthetic flows or LES data, since the dynamics of all Fourier modes, i.e., the
full turbulent energy spectrum, is resolved numerically, and the filter is
applied a posteriori.
The resulting fields can be considered \emph{perfect}, representing model-free
LES.
However, we point out two differences compared to conventional LES: Firstly, we
resolve the flow field on the original mesh with $N_g=1024^3$ grid points, even
though $N_g=32^3$ nodes per dimension would suffice to resolve the truncated
spectrum in physical space without significant loss of information.
Essentially, this is spectral interpolation, which is avoided in LES in
practice for obvious reasons.
Secondly, the filtered flow field does not exhibit a dissipative range, since
the filter sharply cuts the energy spectrum in the inertial range.
Usually, SGS models, such as the Smagorinsky model, are used in LES.
These models create an artificial dissipative range by increasing the effective
viscosity~\cite{popeTurbulentFlows2000}.

Now, what is the role of the filter?
The direct effect of the filter is that $\u$ and $\A$ in
\cref{eqn:pp,eqn:up,eqn:cons,eqn:xi,eqn:Z} are replaced by the filtered fields
$\bar{\u}$ and $\bar{\A}$, respectively.
These fluctuate on slower in time and on coarser length scales.
While $\tau_\eta$ is increased by a factor of $5$, the advection timescale
$\tau_a$ roughly only doubles, meaning that the Kubo number increases from
$Ku^{\mathsf{(DNS)}}=10.3$ to $Ku^{\mathsf{(FDNS)}}=21.4$, shifting the system
towards the persistent regime.
In the context of particle-laden turbulence, it is important to emphasize that
gradient self-amplification persists in the filtered field
velocity~\cite{johnsonMultiscaleVelocityGradients2024}.
This is verified by the plot in \cref{fig:QR-eul} based on our dataset of
sharp-spectrally filtered DNS, where the characteristic sheared teardrop shape
of the joint probability density of $Q$ and $R$ is recovered.

\section{Results}\label{sec:results}
\subsection{Instantaneous clustering}\label{subsec:instant}
The particles are initialized at uniformly distributed random positions with
the local flow velocity.
All cases start from identical initial conditions at the same instance in time.
Since the particles do not have any memory when they are introduced into the
flow, they cluster in the beginning solely due to the centrifuge mechanism.
This initialization facilitates the treatment of the particle velocity as a
continuum, a 'flow of particles'\cite{leeIdentificationParticleCollision2023},
in the sense that \cref{eqn:up,eqn:cons,eqn:xi} are valid until a caustic
occurs.
Later on, the particle velocity field becomes locally multivalued, and the
Eulerian equations must be abandoned in favor of \cref{eqn:Z}.

\begin{figure}
	\centering
	\includegraphics[width=\columnwidth]{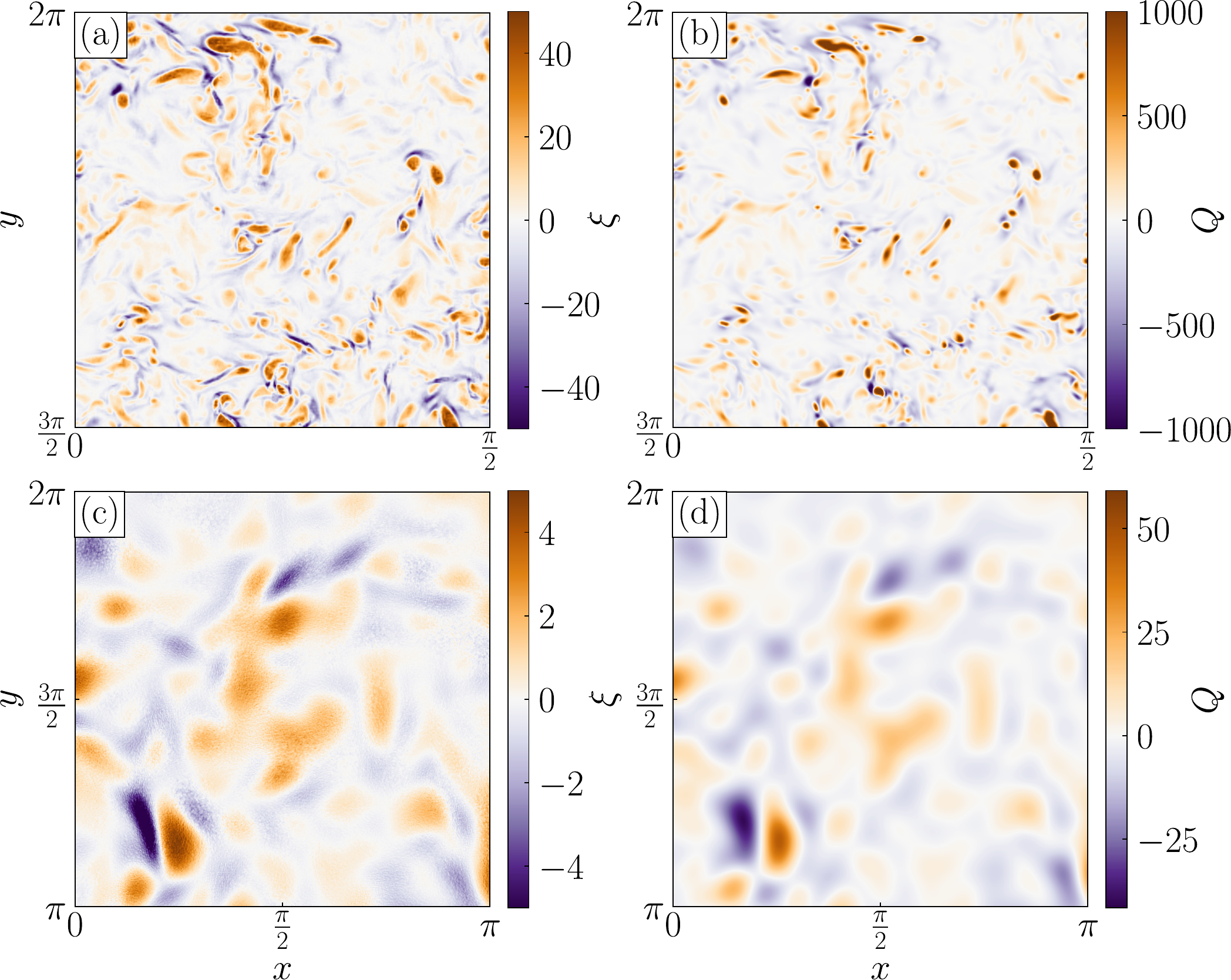}
	\caption{\label{fig:div}
		Instantaneous clustering at short integration times.
		Grid-projected particle velocity divergence fields of the (a)~\DNSB~and
		(b)~\LESB~cases at time $t=0.04$ and $t=0.08$, respectively.
		In panels (b) and (d), contours of the local instantaneous $Q$ fields are
		shown.
		The contours of panel (b) are clipped to the range $-1000\leq Q\leq1000$, in
		order to hide the rare extreme excursion of the unfiltered field in the color
		map, whereas the $\xi$-fields in the panels (a) and (c) are clipped to
		$-50\leq\xi\leq50$ and $-5\leq\xi\leq5$, respectively.
	}
\end{figure}
\Cref{fig:div} shows the divergence of the particle velocity field alongside contours
of the local instantaneous $Q$ field at an early instance in time, at which particles have not yet acquired a significant amount of memory, and only very few particles paths have crossed.
First, notice how the sharp spectral filter removes the smallest scales of the
flow field and thereby smoothens the velocity gradients, as well as massively
narrows the range that $Q$ attains.
The color map in \cref{fig:div}(b) is clipped at $Q=\pm1000$,
preventing the contours from being washed out by extreme outliers.
Nevertheless, intermittent small-scale regions of intense turbulence are
observed, where $Q$, in reality, attains values of up to $10^5$.
This reflects the nature of turbulence, as the smallest scales exhibit the most
extreme vorticity and strain, produced by the non-linear self-amplification of
the Navier-Stokes equation.
Second, notice the excellent visual agreement between the contours of the
divergence patterns (Figs.\ref{fig:div}(a,c)) and
$Q$ (Figs.\ref{fig:div}(b,d)).
'Particle sinks' with positive values of $\xi$ emerge in regions where $Q$ is large and positive, i.e., where particles are ejected out of vortices.
Particles accumulate, on the other hand, where $\xi$ is negative, leading to an
increase of the particle density via \cref{eqn:cons}.
These regions coincide with large negative values of $Q$, i.e., where strain
overwhelms vorticity.
In other words, the instantaneous flow field explains the spatial distribution
of the particles well at this early stage, and Maxey's centrifuge gives
qualitatively accurate predictions of clustering.

\begin{figure*}
	\includegraphics[width=\textwidth]{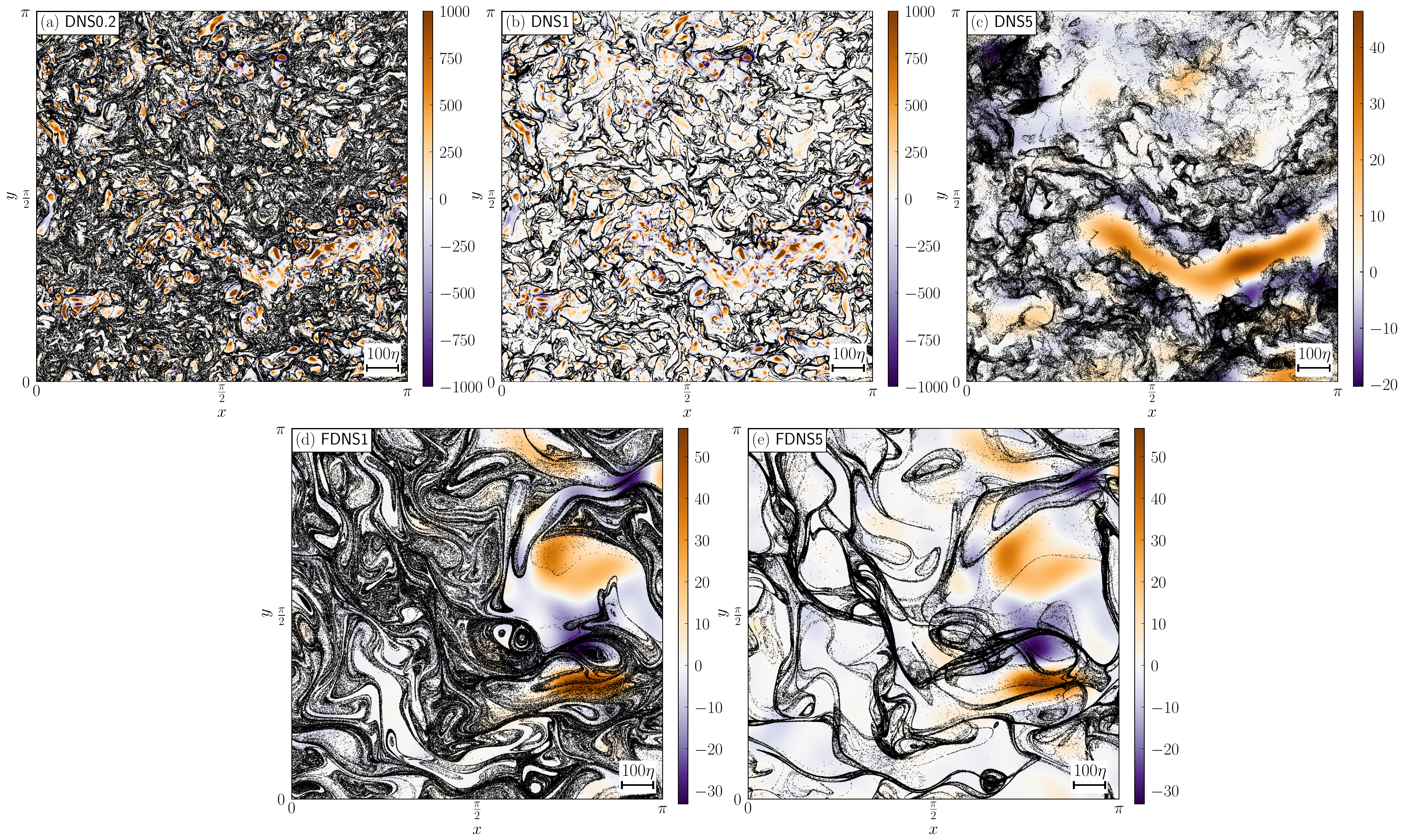}
	\caption{\label{fig:poc}
		Spatial distribution of particles at steady state, with contours of $Q$ in the
		background.
		The particles inside a slice of thickness $s=\eta$ are represented by black
		dots.
		Only a quarter of the domain is shown.
		As in \cref{fig:div}, the range of $Q$ is clipped in panels (a) and (b).
		Note that the background of panel (c) is obtained from the filtered flow field,
		even though the particles are suspended in the full-spectrum unfiltered flow.
	}
\end{figure*}

\Cref{fig:poc} visualizes the particles in thin slices
of thickness $s=\eta$ after long integration time with the contours of local
instantaneous $Q$ in the background.
The particles shown have reached steady state, both in terms of the density
variance $\langle(n-\langle n\rangle)^2\rangle$ and its anti-correlation to
$Q$.
The sharp spectral filter was applied to the flow in the background of the
\DNSB~case displayed in \cref{fig:poc}(c) in order to remove turbulent
eddies with timescales $\tau<\tau_p$ from the flow; in the simulation, however,
the unfiltered flow was used.

While \cref{fig:div} showed how the granularity of the flow is imprinted on the
particle fields after initializing the simulations, \cref{fig:poc} illustrates
that the coarseness of the particle distributions persists even after long
integration times.
We remark that additional simulations were performed in which the particles
were initialized with randomly sampled isotropic velocities.
The imposition of the particles' granularity by the flow field was also
observed in that case (not shown).
In fact, the patterns that emerged after long integration times showed little
to no differences to those observed in \cref{fig:poc}.
The persistence of the particle field granularity implies that SGS velocity
fluctuations are needed in LES to recover qualitatively accurate spatial
distributions of inertial particles.

Clearly, particle clustering is observed in all cases.
In the \DNSA~case with $St=1$, which is known to exhibit the most intense
preferential concentration, clusters are thin, curved structures that are
elongated along the local flow topology.
Indeed, \cref{eqn:maxey} seems to predict the spatial distribution of the
particles qualitatively well in this case, even after long integration times.
As expected~\cite{becStatisticalModelsDynamics2024a}, these patterns match the
typical range of preferential concentration, $\ell\sim10\eta$, with void sizes
spanning into the inertial range of turbulence.
A prominent large gap in the particle fields (located in the region
$(\tfrac{\pi}{2}\leq x\leq\pi,0\leq y\leq\tfrac{\pi}{2}$)) is seen in all
unfiltered cases, coinciding with the largest connected region of increased $Q$
of the coarse field (\cref{fig:poc}(c)).
Similarly to the \DNSA~case, the \DNSC~particles ($St=0.2$) avoid high-$Q$ and
populate low-$Q$ regions, respectively.
Particle voids are, however, more exclusive to the most intense vortices.
In-between clusters, the particles are distributed more uniformly, and
small-scale filamentary patterns are observed.
The path-history effect produces fractal clusters with a decrease in the
fractal dimension as the Stokes number is increased from $St=0.2$ to $St=1$.
In the \DNSB~case ($St=5$), the particles are too heavy to resonate with the
smallest scales of the flow field, which act to randomize the particle fields
and impose their granularity.
Once the fine scales are removed, however, qualitative agreement between
clusters (voids) and low (high) $Q$ regions is observed, as seen in
\cref{fig:poc}(c).
Clustering agrees best with the most significant coarse-grained flow contours
that match the particle timescale, $\tau_p$.
Note that the background in \cref{fig:poc}(c) was obtained by filtering the
flow field explicitly to match $\tau_p$, i.e., so that only turbulent
structures of timescale $\tau\leq\tau_p$ are removed.
This provides evidence for the self-similarity of particle clusters, previously
reported by Goto and Vassilicos~\cite{gotoSelfsimilarClusteringInertial2006}.
The authors' reasoning is based on the premise that turbulence consists of a
hierarchy of self-similar eddies across different length scales, each
resonating with particles of a specific $\tau_p$.

In the filtered cases (Figs.~\ref{fig:poc}(d,e)),
instantaneous preferential sampling does not explain clustering well.
Both cases exhibit significantly elevated particle number density in low and
high $Q$ regions.
Moreover, the spatial distributions reveal intricate structures with
filamentary patterns.
The particle density fields are highly intermittent, with sharp edges
separating clusters from voids.
The sharp edges are seen because the number of particles remains unchanged
between the unfiltered and filtered cases, yet the resolution is higher
relative to the smallest scales.
Filaments are most conspicuous in the \LESA~case, where particles fill the
domain more uniformly; however, they are also seen in~\LESB, where caustics
contribute to the formation of highly irregular patterns.
Small-scale filamentary clusters that do not coincide with the local patterns
of strain or vorticity have been observed before (e.g. in Fig.
1(b) in Ref.~\onlinecite{gustavssonStatisticalModelsSpatial2016}),
yet the strong misalignment of particle density and $Q$ as well as the large length scale
at which filaments are seen are surprising.
A possible explanation for the discrepancy between the effectiveness of the
centrifuge mechanism to explain clustering in the unfiltered and filtered cases
is that the filter removes the regions of the most extreme vorticity and
strain, inherent to the dissipative scale of turbulence.
We are not convinced by this explanation, however, as, in the smallest scales,
the particle fields of the unfiltered cases reveal similar patterns that cannot
be attributed to the instantaneous flow field.
This is shown in the following section.

\subsection{Scale similarity} \label{subsec:similarity}
\begin{figure*}
	\includegraphics[width=\textwidth]{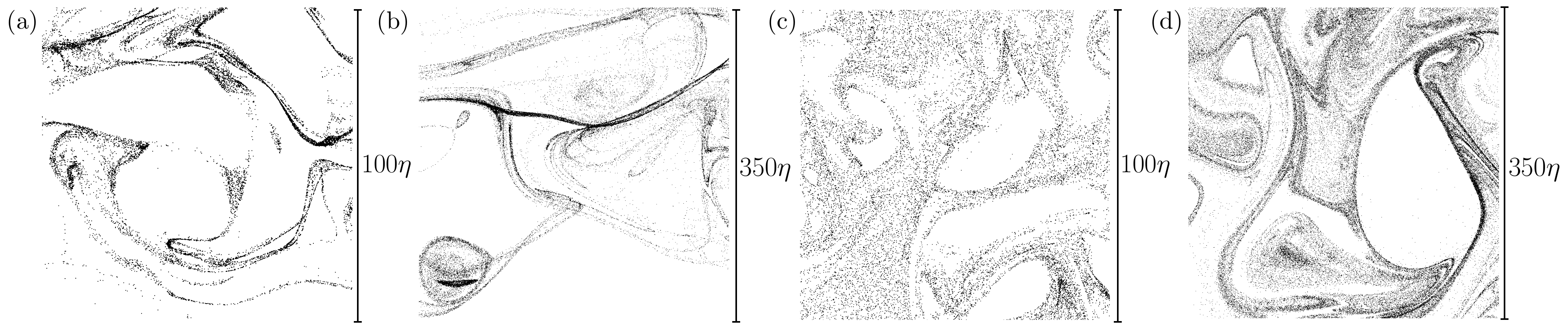}
	\caption{\label{fig:similarity}
		Scale-similarity of the particle fields from the unfiltered and filtered at
		matching (effective) Stokes numbers.
		Particles in a slice of thickness $s=\eta$ are shown as black dots.
		(a) \DNSA~with $St=1$, (b) \LESB~with
		$\Steff=0.98$, (c) \DNSC~with $St=0.2$, and (d) \LESA~with $\Steff=0.2$.
	}
\end{figure*}
Filtering out small-scale turbulent motions leads to an increase in the
characteristic flow timescale.
For the filtered fields, this timescale is calculated as\cite{gustavssonStatisticalModelsSpatial2016}
\begin{equation}\label{eqn:taueta}
	\bar{\tau}_\eta = \big\langle\tr(\bar{\A}\bar{\A}^\mathsf{T})\big\rangle^{-1/2}.
\end{equation}
(Note that, for the sake of simplicity, the symbol $\tau_\eta$ was used without an overbar in \cref{tab:table1} for all cases.)
We then define the filter-effective Stokes number as
$\Steff=\tau_p/\bar{\tau}_\eta$.
With the Stokes numbers adjusted to the timescale of the filtered flow, we
expect to observe similar fractal properties (i.e., matching correlation
dimensions) and spatial patterns between the~\DNSC~and~\LESA, as well as
between the \DNSA~and~\LESB~cases, respectively.

Now, at what length scale should the spatial patterns of the unfiltered and
filtered cases be compared?
Here, it is important to recall that the filtered fields are obtained by
sharply truncating the Fourier spectrum at a wavenumber in the inertial range,
$\k_c=16$.
The filtered fields therefore do not exhibit a dissipative range.
Even though the subscript $\eta$ is used for the filtered flow timescale
(\cref{eqn:taueta}), it may not be thought of as the 'Kolmogorov' timescale.
In fact, the notion of the Kolmogorov range of turbulence is not applicable to
the filtered fields.
While \cref{eqn:taueta} is a valid expression for the characteristic timescale,
the identification of the reference \emph{length} scale is more complicated.
Ray and Collins~\cite{rayPreferentialConcentrationRelative2011} use the
filtered energy spectrum to compute the dissipation as $\bar{\epsilon} =
	2\nu\int_{0}^{\k_{c}} \k^2 E(\k)\, d\k$, where $E(\k)$ is the turbulent kinetic
energy associated with the Fourier mode $|\bm{\k}|=\k$.
The reference length scale of the filtered fields was then calculated as
$\bar{\eta}=(\nu^3/\bar{\epsilon})^{1/4}$.
This quantity is, however, elusive in our setup, as nearly all of the
dissipation of turbulent kinetic energy occurs in the dissipation range, i.e.,
at the largest wavenumbers.

Here, we introduce an ad hoc method to estimate the length at which the
unfiltered and filtered cases should be compared: We compute the ratio between
the largest wavenumbers arising in the inertial range of the turbulent flow in
the filtered and the unfiltered flow fields, that is the filter cut-off and the
end of the inertial range, $\k_{DI}/\k_c\approx3.5$.
The 1-to-3.5 scaling is employed in \cref{fig:similarity}, which shows slices
of particles of the cases~\DNSA,~\LESB,~\DNSC~and~\LESA.
The length scales of clusters and voids are comparable.
We are again confronted with filamentary patterns observed earlier in the
spatial distributions shown in \cref{fig:poc}.
In addition, there is a noticeable decrease in the extent to which particles
fill the domain as the (effective) Stokes number increases, suggesting a
reduction in fractal dimension.

Scale similarity between the \DNSA~and~\LESB~as well as between the
\DNSC~and~\LESA~cases is visually evident, i.e., similar patterns are observed
when magnified to the appropriate level.
The \DNSA~clusters are revealed to have internal structure which where
previously (\cref{fig:poc}(b)) not visible.
These patterns are indicative of multiplicative amplification, which, in
simulations of statistical models\cite{gustavssonStatisticalModelsSpatial2016},
is observed at length scales even smaller than $\eta$.
Numerical studies comparable to the present work usually do not resolve such
structures and thus obtain particle fields that can be explained well at
sufficiently large scales based on the instantaneous flow fields, not unlike
panels (a) or (b) in \cref{fig:poc}.
Some examples are Fig.~2(a) in
Ref.~\onlinecite{becStatisticalModelsDynamics2024a}, Fig.~2 in
Ref.~\onlinecite{cenciniDynamicsStatisticsHeavy2006}, or Fig.~5 in
Ref.~\onlinecite{wangReynoldsNumberDependence2020}.
Based on the visual evidence provided by \ref{fig:similarity}(a), however, it
must be expected that small-scale filaments are present throughout, yet usually
not visible due to the level of magnification or unresolved due to an
insufficient number of particles.
These filaments are the result of multiplicative amplification, and their
arrangement does not align well with the instantaneous flow field.

\begin{table}
	\caption{\label{tab:corr}
		Correlation dimensions.
	}
	\begin{ruledtabular}
		\begin{tabular}{c c c c}
			Case &$St$&$\Steff$&$d_2$\\ \DNSC & 0.20 & - & $2.64$\\ \LESA & - & 0.20 &
			$2.73$\\ \DNSA & 1.00 & - & $2.33$\\ \LESB & - & 0.98 & $2.41$\\ \end{tabular}
	\end{ruledtabular} \end{table} In line with
Ref.~\onlinecite{becMultifractalConcentrationsInertial2005}, the fractal
dimension of the particle distributions decreases with increasing Stokes
number.
Measurements of the correlation dimensions are provided in \cref{tab:corr}.
The values were obtained by performing polynomial fits over the radii $r/\eta
	\in \{0.5,1,1.5,2\}$ and averaging over ensembles of 75 randomly selected test
particles for each case.
The $St$-dependence of the correlation dimension matches the values from the
literature, recently collected by
Bec~\etal~\cite{becStatisticalModelsDynamics2024a} The effective Stokes number
therefore approximately predicts the fractal quality of the particle clusters.
We also find that unfiltered and filtered cases with comparable (effective)
Stokes numbers yield similar correlation dimensions, confirming spatial
scale-similarity quantitatively.

\subsection{Sling events}\label{subsec:sling}
We now turn to the discussion of sling events, which occur throughout our
simulations.
In the following, we provide detailed observations of these events and test the
applicability of new theories for their prediction.
\begin{figure*}
	\centering
	\includegraphics[width=0.75\textwidth]{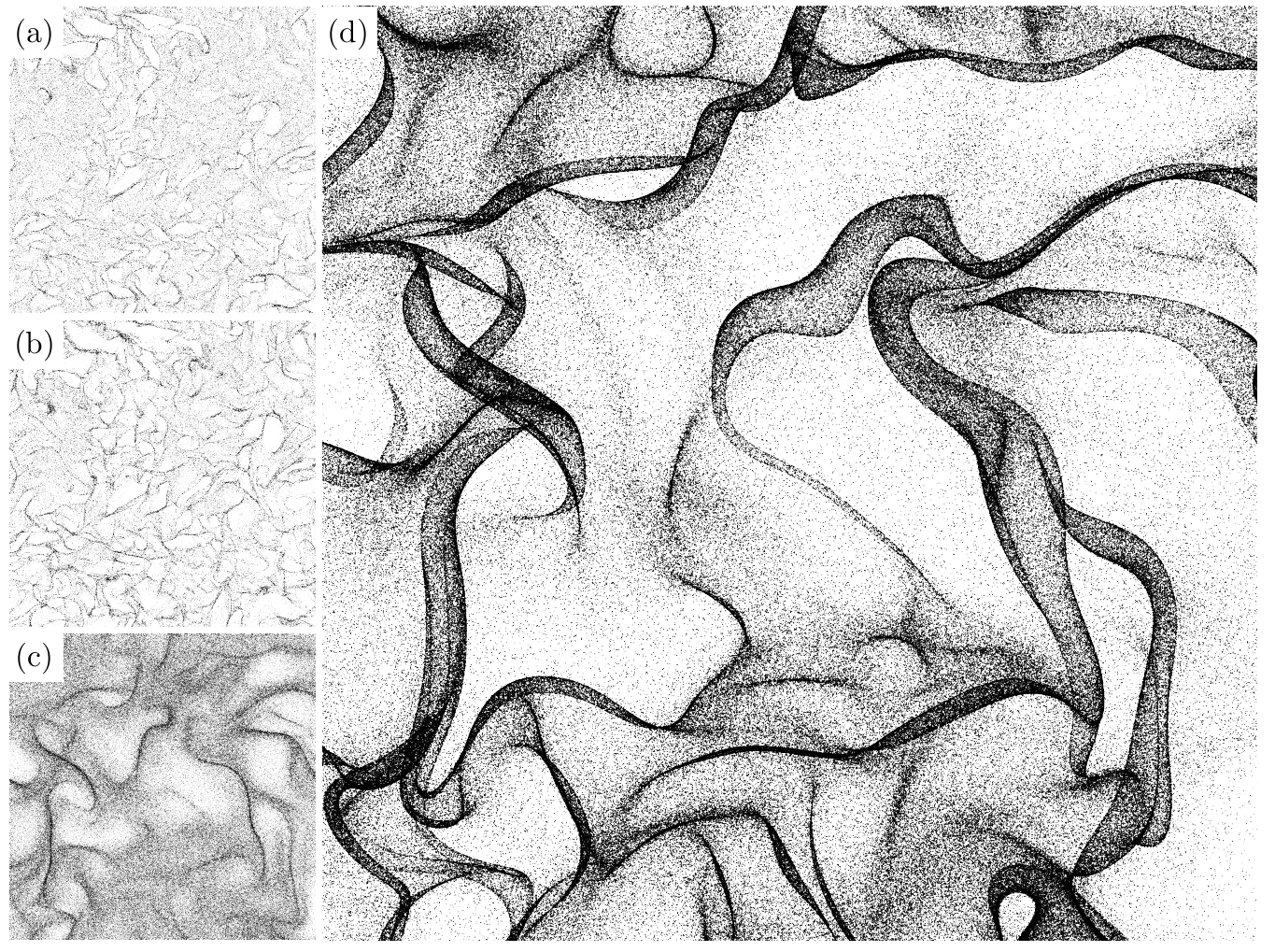}
	\caption{\label{fig:early}
		Caustic patterns in unfiltered ((a)~\DNSA,~(b)~\DNSB) and filtered
		((c)~\LESA,~(d)~\LESB) turbulence at times (a,b) $t=0.012\approx3.8\tau_\eta$
		and (c,d) $t=0.8\approx2.8\bar{\tau}_{\eta}$.
		The particles in a slice of thickness $s=\eta$ are depicted as black dots.
		Panels (a,b) show an eighth of the domain, whereas (c,d) show a quarter.
	}
\end{figure*}
Patterns of caustics are observed in all cases, as seen in \cref{fig:early},
which shows particles scattering in the computational domain after short
integration times.
We emphasize that modeling the particle motion using an Eulerian continuum
formalism (\cref{eqn:up}) is clearly not suitable under these circumstances, 
as significant numbers of particles undergo caustics and
therefore attain multivalued velocities at the same location in space and time
(see one-dimensional model presented in \cref{sec:intro}).
In the filtered cases, caustics are conspicuous thanks to their
warped manifold structure at comparatively large length scales,
populated by large ensembles of particles.
\Cref{fig:early}(d) visualizes the peculiar patterns that are associated
with sling events at large Stokes number.
Heavy particles exhibit high relative velocities and form
'pockets'~\cite{bewleyObservationSlingEffect2013} (triggering collisions in
reality), whereas lighter particles have lower memory capacity and adjust to
the flow easily, meaning that lower relative velocities are attained during
caustics.
This explains the absence of pockets in the \LESA~case, where caustics appear
as lines (sheets in 3D).
The pockets in the \LESB~case are framed by sharp lines of high particle number
density, which will be shown to exhibit maximal relative velocities
(\cref{fig:les5caustic}).
A 3D visualization of the pocket character of sling events in the \LESB~case is
provided as supplementary material in \cref{fig:3dbacktrack}(b).

Now, how do we classify the particles that participate in caustics?
Meibohm~\etal~\cite{meibohmPathsCausticFormation2021,meibohmCausticsTurbulentAerosols2023,meibohmCausticFormationNonGaussian2024}
integrate \cref{eqn:Z} alongside \cref{eqn:pp} for each particle and identify
sling particles as those that reach $\tr(\Z)$ below a critical value.
Similarly, Lee \& Lee~\cite{leeIdentificationParticleCollision2023} use an
Eulerian model for the particle motion and identify caustics as blowups of
\cref{eqn:xi}.
Both methods are meaningful in that they provide the exact instance in time at
which the caustic occurs, yet they are computationally inconvenient.
In this work, a simpler, computationally cheaper method is used to condition
the particles.
Our approach exploits the key characteristic of the caustic: multivalued
particle velocities, i.e., close-by particle pairs $(l,k)$ having large
relative velocities $\mathrm{d}v_{lk}=|\v_l-\v_k|$.
For each particle, we compute $D=\langle\mathrm{d}v\rangle_{3nn}$, where
$\langle\cdot\rangle_{3nn}$ is the ensemble average over the 3 nearest
neighbors, which are only accounted for if all three particles are not further
away than $\eta$.
Particles that do not have at least three neighboring particles inside the
sphere of radius $r=\eta$ are exempted.
We then pick a suitable instance in time, $t_c$, and identify particles with
$D$ exceeding a certain threshold, $D_c$.

Our method is computationally convenient and can be applied to any numerical
simulation that uses particle tracking, without requiring the integration of
additional equations.
No extra fields need to be tracked, and the FVG tensor only needs to be
computed at time $t_c$, eliminating the need to calculate and store both $\A$
and $\Z$ for each particle at every time step.
However, we acknowledge that, unlike methods that explicitly track blow-ups in
\cref{eqn:Z}, our method is not able to capture the \emph{exact} instance in
time when the caustic occurs.
When computing the instantaneous relative velocities, there is some uncertainty
associated in whether a particle has just reached a large value of $D$, i.e.,
if the caustic has \emph{just} occurred, or whether the particle is currently
in the onset of a caustic and the main jump in $D$ is yet to take place.
A preliminary study, however, indicated that particles tend to undergo swift,
steep increases in $D$, rather than smooth ascents over longer periods of time.
Uncertainty in the peaks of $D$ is especially limited in the unfiltered cases,
where the spikes take place over the course of only a few time steps.
\begin{figure*}
	\includegraphics[width=\textwidth]{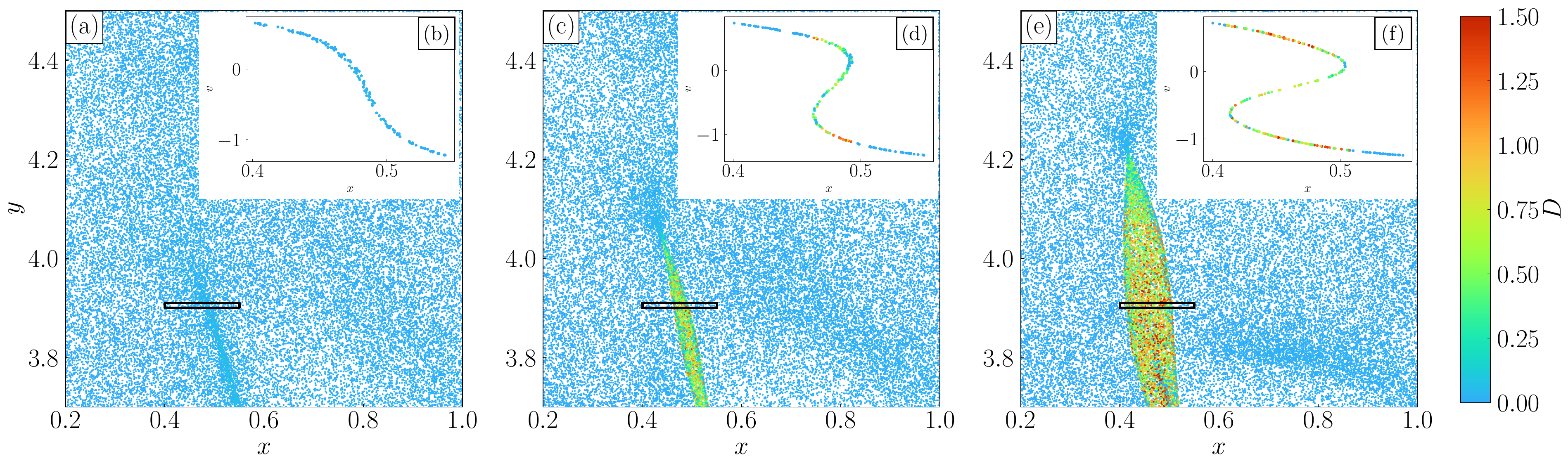}
	\caption{\label{fig:les5caustic}
		Identification of a caustic in a sling event, obtained in the \LESB~case after
		(a,b) $t=0.14$, (c,d) $t=0.2$ and (e,f) $t=0.28$.
		Panels (a,c,e) show the particles scatter in a slice of thickness $s=\eta$,
		colored in the ensemble-averaged relative velocity $D$, while the insets
		(b,d,f) depict the $v$-$x$ phase-space of the particles inside the indicated
		region.
	}
\end{figure*}

As described in \cref{sec:dns}, \cref{eqn:pp} is solved in three dimensions for
a large ensemble of particles that lag the turbulent carrier gas flow.
We find that caustics form readily, and as early as after $t\approx2\tau_\eta$.
To provide an example, we visualize the identification of a sling event in the
\LESB~case in \cref{fig:les5caustic}.
The large panels depict the particles scattering in a thin slice at three
instances in time, colored based on the value of $D$.
A slender area of increased relative velocity is easily identified.
The sling region, initially a thin line in \cref{fig:les5caustic} (a) (implying a
sheet in 3D), attracts particles and grows in width to an elongated cusp.
This indicates that, independent of Stokes number, sling events occur 
on thin sheets, and that pockets (as seen in \cref{fig:early}(d)) develop later in time at large enough $St$.
Particles with opposing velocities approach each other from the negative and positive $x$-directions and cluster in the region around $x\approx0.5$.
Here, the particles are focused into a narrow region and the particle number
density increases locally by a factor of around $2.5$ (not shown).
The transition from a smooth, unambiguous to a multivalued particle velocity
field is evident from the insets in \cref{fig:les5caustic}, showing the phase
space composed of the first velocity and position components of the particles
inside the small volume spanning across the caustic and marked by the black
rectangle.
The signal folds over itself, not unlike the one-dimensional model problem seen
in \cref{fig:1dcaustics}.
The insets also reveal that the largest relative velocities exist towards the
edge of the caustic manifold, where 'new', freshly supplied particles are
introduced into the sling area.
In this region, particle collisions would occur most violently, as particles
undergo a sharp increase in relative velocity.

\begin{figure*}
	\includegraphics[width=\textwidth]{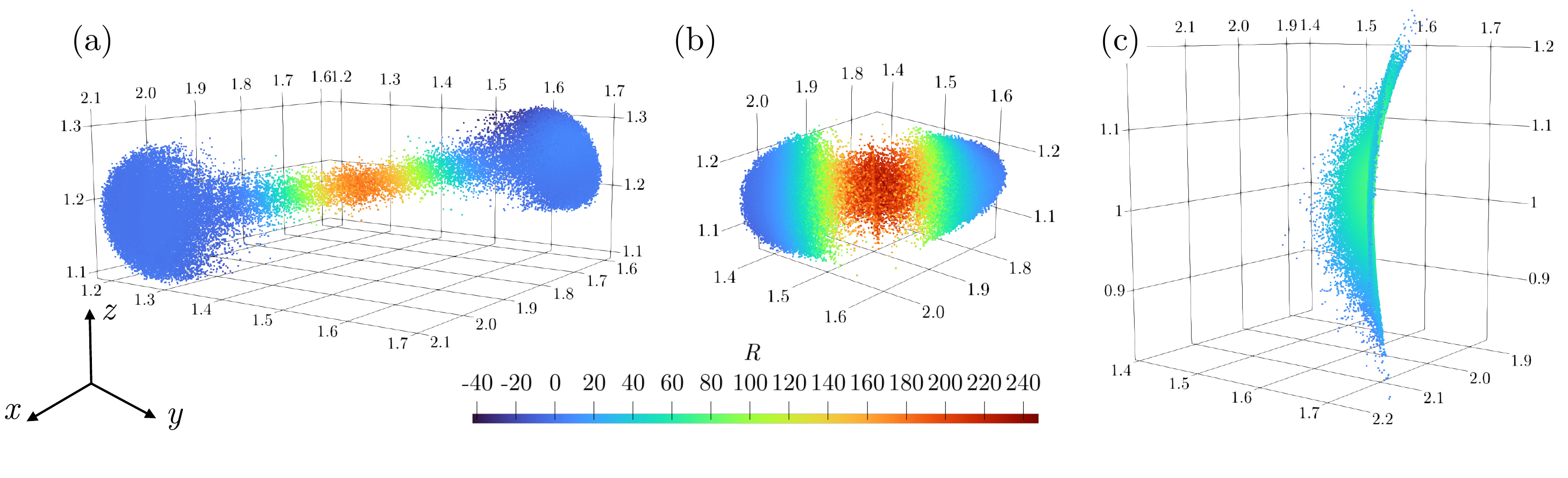}
	\caption{\label{fig:3d}
		3D renderings of a sling event from the \LESA~case at times
		(a) $t=t_c-0.4$, (b) $t=t_c-0.2$
		and (c) $t=t_c$, colored in the locally experienced $R$.
	}
\end{figure*}
In \cref{fig:les5caustic}, we have illustrated the identification of a sling
event by conditioning particles on the nearest-neighbor averaged relative
velocity, $D$, and demonstrated how phase-space folds occur in these regions.
To build on this, we now depict the flow conditions experienced by particles
leading up to the occurrence of a caustic.
For the first time to the best of our knowledge, we present 3D renderings of a
sling event in \cref{fig:3d}.
The visualizations are derived from the \LESA~case, with individual particles
represented as points at three consecutive time instances, colored in the
locally experienced value of $R$.
As described above, the ensemble of sling particles is selected at time
$t_c=0.6$ using the condition $D(t=t_c)>D_C$.
The time instance $t=t_c$ is shown in panel (c), clearly
revealing that the sling event occurs on a thin, warped sheet in 3D space.
The selected ensemble is then tracked backward in time, and the local flow
conditions are interpolated to the particle positions to obtain
Figs.~\ref{fig:3d}(a,b) at times $t=0.2$ and $t=0.4$,
respectively.
An additional 3D visualization of the backtracking of sling particle
trajectories is provided in the appendix (\cref{fig:3dbacktrack}).

Long before the caustic occurs, the 'epicenter' of the sling event can be
identified in \cref{fig:3d}(a).
This area features elevated $R$, but particles have not clustered or collided
yet.
In the following, the particles inside a dumbbell-like volume approach the
epicenter, in which, under the compressive straining of the flow, particles
converge on a thin, compact sheet.
At time $t=t_c-0.2$ (\cref{fig:3d}(b)), the 'sling sheet' commences to take
shape, and it is likely that particles in this region already experience
elevated relative velocities.
Eventually, all particles coalesce into a smooth manifold whose thickness is of
the order of the particle diameter (\cref{fig:3d}(c)).
The fact that we see the sling event taking place on thin sheets is due to the
flow topology in the fourth quadrant of the $QR$-plane, where straining acts
compressive in one and expanding in the other two directions.
Owing to the low effective Stokes number $\Steff=0.2$, however, the particles
collide with comparatively low relative velocities, and $D$ decreases rapidly
after $t=t_c$, as the particles relax to the local flow field and coalesce back
into a smooth flow of particles.
In the \LESB~case, on the other hand, particles collide at much higher relative
velocities and the sheet structure is not retained later in time (see also
\cref{fig:3dbacktrack}(b)).
The reason is that the heavier particles cannot adapt to changes of the flow
velocity swiftly.
This results in higher relative velocities at the time of the sling event and
longer times during which they are retained.
The crossing of particle paths is responsible for the development of pockets.

\begin{figure}
	\includegraphics[width=\columnwidth]{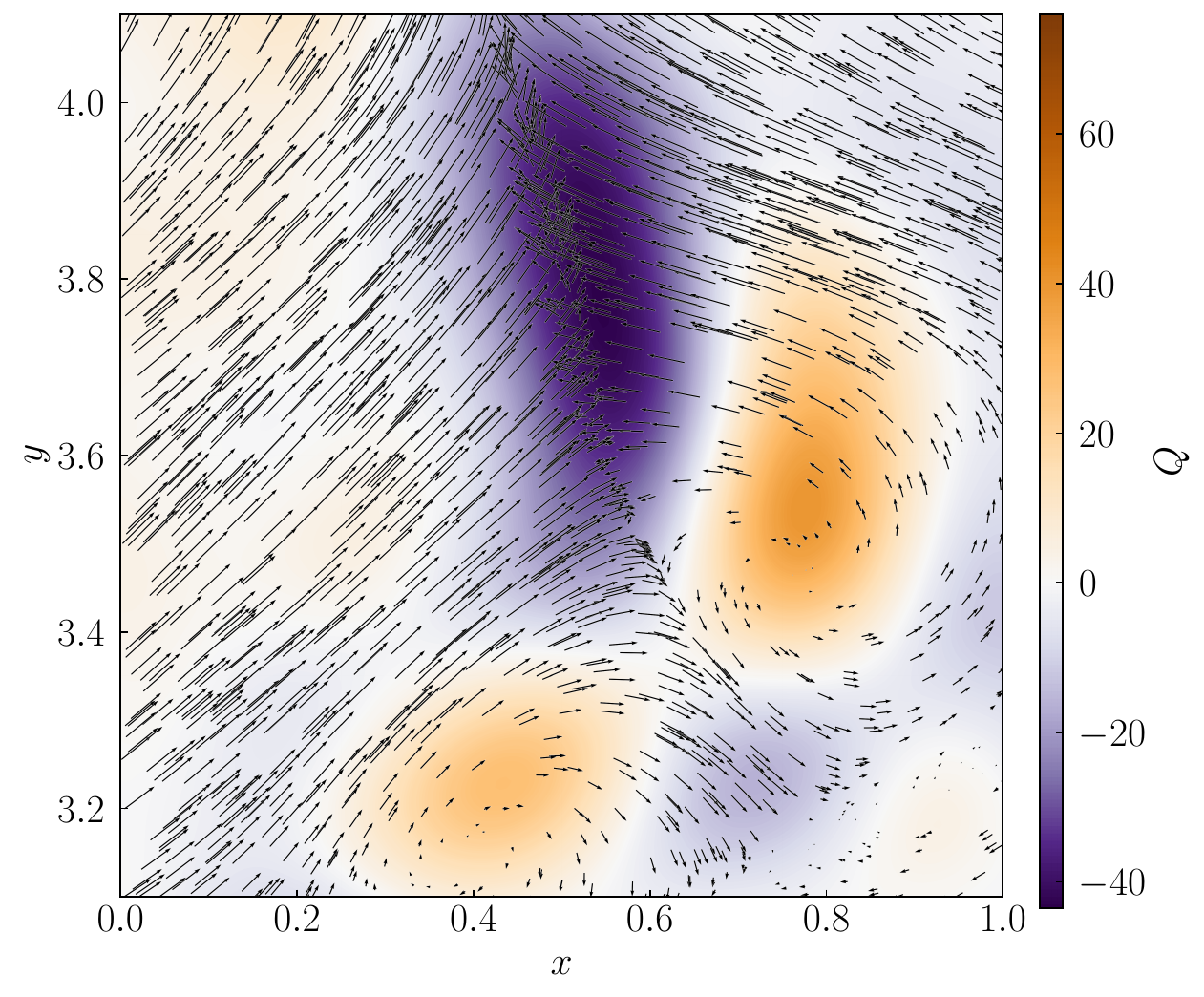}
	\caption{\label{fig:quiver}
		Onset of a sling event in the \LESB~case at $t-t_c=0.2$.
		The $x$-$y$ velocities of the particles in a slice of thickness $s=0.03\eta$
		are depicted as arrows, whereas the contours of local $Q$ are shown in the
		background.
	}
\end{figure}
While the 3D renderings of \cref{fig:3d} only depicts particles that actually 
participate in the sling event, \cref{fig:quiver} provides a 2D view of particles
for which no particular selection was applied.
The figure shows particles of the \LESB~case at time $t=t_c+0.2$ (same time
instance as in Figs.~\ref{fig:les5caustic}(c,d))
as arrows pointing in the direction of the $x,y$-velocities.
The contours $Q$ are shown in the
background and delineate two vortices, separated by a region of intense strain
around $x=0.5$.
The flow directions of the particles visualize the sling event that takes place
in the $Q<0$ region.
Here, the particle velocities are multivalued due to the occurrence of the
caustic, as observed in \cref{fig:les5caustic}(d).

\Cref{fig:quiver} also depicts the $Q$-conditions that particles
experience before participating in the sling event.
The flow directions indicate that particles are centrifuged out of the vortical
structures and then engage in sling dynamics.
However, not all particles that escape the $Q>0$ regions will participate in
the sling event.
For examples, those that are transported in the negative $y$-direction do not
attain multivalued velocities.
This indicates that, upon being ejected out of the vortices, a prolonged
residence in flow topologies of compressive strain is required for a particle to participate in a sling event.
This is the case for flow topologies that progress along positive branch of the
Villeifosse tail.
\Cref{fig:quiver} is consistent with the finding by Lee \&
Lee~\cite{leeIdentificationParticleCollision2023} that caustics occur in sheets
'packed' in between vortex layers.
Particles are ejected out of vortices from opposite sides and, under
compression strain from the turbulent motion, collide in between, where the
curl is low and strain dominates.
This insight is not contradictory to the traditional picture of preferential
concentration, however the centrifuge mechanism appears to merely act as
preconditioner for caustics.

\begin{figure*}
	\includegraphics[width=\textwidth]{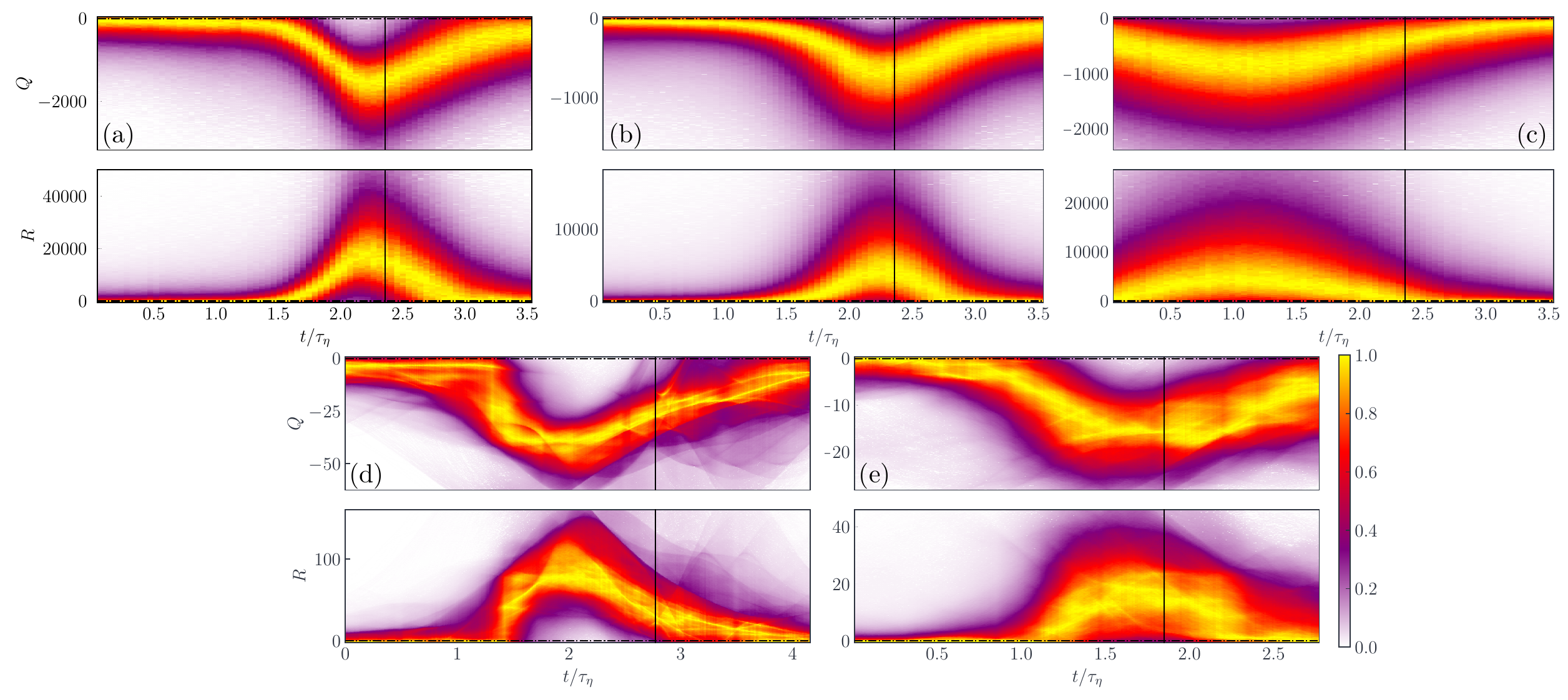}
	\caption{\label{fig:tracking}
		Time evolution of the densities of $Q$ and $R$ along sling particle
		trajectories for the cases (a)~\DNSC, (b)~\DNSA, (c)~\DNSB, (d)~\LESA,
		(e)~\LESB.
		At each time step, the trajectory density is scaled by the maximum value.
		The black vertical lines depict the time $t_c$ at which the sling particles are
		selected.
	}
\end{figure*}

The visual observations from \cref{fig:3d,fig:quiver} point to the fact that
particles undergoing caustics experience intense turbulent motion in which they
sample FVGs heterogeneously.
As a matter of fact, throughout our simulations, sling particles move through
flow topologies of large negative $Q$ and large positive $R$.
This statistical bias away from $\langle Q\rangle=\langle R\rangle=0$ is
clearly evident in \cref{fig:tracking}, which visualizes the most likely paths
that sling particles take in terms of the FVG tensor invariants.
The figure was produced as follows: For each case, we fix the threshold $D_c$
and identify an ensemble of sling particles at time $t=t_c$ (marked by the
vertical black lines in \cref{fig:tracking}) using the condition
$D(t=t_c)>D_c$.
These particles are then tracked backward in time (see \cref{fig:3dbacktrack}
for a supplementary visualization) and each point along the particle
trajectories, the probability densities $p(Q)$ and $p(R)$ of the ensembles are
computed.
The threshold $D_c$ was adjusted slightly between the different cases in order
to ensure similar numbers of tracked particles, which is necessary since the
particles collide at vastly different relative velocities, depending on the
Stokes number.
It was, however, ensured that the choice of $D_c$ in the ranges considered has
only minor influences on the trajectory densities shown in \cref{fig:tracking}.
We found that the higher we chose $D_c$, the more extreme excursions of $Q$ and
$R$ particles experience, however the curves of the fluctuations remain
identical in shape.
We note that alternative methods that track blowups of
\cref{eqn:Z}\cite{meibohmCausticsTurbulentAerosols2023,meibohmCausticFormationNonGaussian2024}
or \cref{eqn:xi}\cite{leeIdentificationParticleCollision2023} are not exempted
from the choice of an ad-hoc threshold, and we expect that the distributions
over ensembles of sling particles of $\tr(\Z)$, computed by explicitly
integrating \cref{eqn:Z}, also varies considerably with the Stokes number.
\begin{figure*}
	\includegraphics[width=\textwidth]{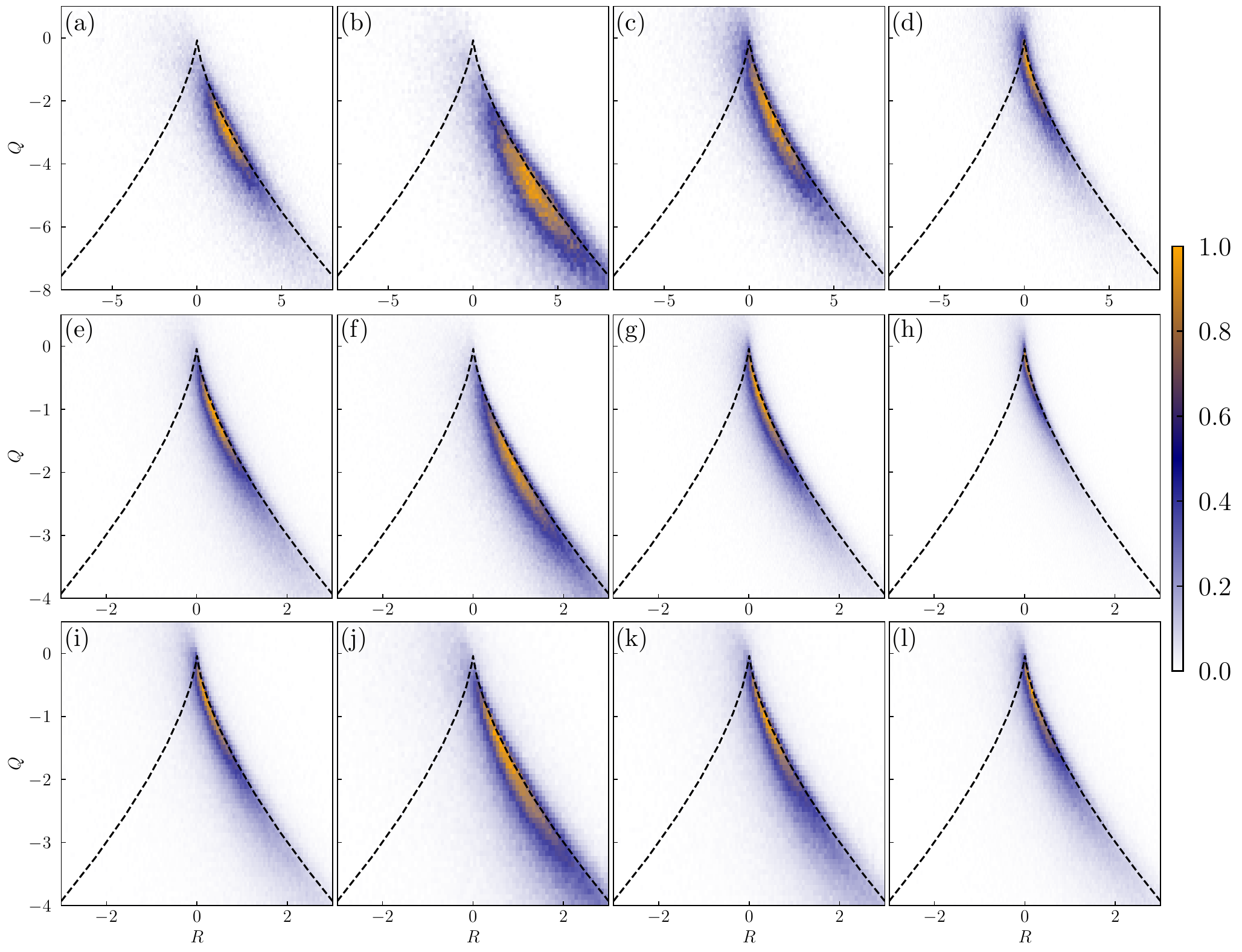}
	\caption{\label{fig:QR}
	Joint probability of $Q$ and $R$ along particles trajectories of the
	(a-d)~\DNSC, (e-h)~\DNSA~and (i-l)~\DNSB~cases, scaled by the maximum value at
	each time step.
	Time series: (a-d) and (e-h) $t=[0.08,0.1,0.12,0.14]$, (i-h)
	$t=[0,0.04,0.08,0.1]$.
	The vertical and horizontal axes are scaled with $A_0$ and $A_0^{-3/2}$,
	respectively, where $A_0$ is an arbitrary constant, (like in
	\cref{fig:QR-eul}), however panels (e-l) show smaller regions of the
	$QR$-plane.
	The dashed black line depicts the Vieillefosse line.
	}
\end{figure*}

First, note how the scales of $Q$ and $R$ along the trajectories differ
significantly between the filtered and unfiltered cases.
In unfiltered turbulence, violent short-lived excursions of $\A$ are obtained,
with isolated particles escaping to values of $R=10^{6}$ and $Q=-10^4$
(hidden by the axes limits of \cref{fig:tracking}).
This is a consequence of the well-known nature of turbulent flows to produce
the most extreme strain and vorticity intermittently in the dissipative
range~\cite{meneveauLagrangianDynamicsModels2011}, offering a possible
explanation to why sling events are observed at higher frequency in the
unfiltered cases.
In clouds, it is believed\cite{fouxonIntermittencyCollisionsFast2022} that high
levels of intermittency (owing to large Reynolds numbers of up to
$\Re_\lambda\sim10^4$) are responsible for the sling dynamics, enhancing
collisions and coalescence of droplets, which initiates rainfall.
The filter decreases the probability of rare extreme events in the flow field.
This reduces the likelihood that particles will encounter the persistent
intense flow topologies necessary to trigger caustics.

Qualitatively, the shapes of the excursions away from $Q=0$ and $R=0$ exhibit
only minimal differences between \DNSC~and~\DNSA.
In fact, both Figs.~\ref{fig:tracking}(a,b) enjoy significant
visual agreement to Fig.~3 in
Ref.~\onlinecite{meibohmCausticsTurbulentAerosols2023}, which was produced by
tracking particles with a Gaussian model at $St=0.3$ and $Ku=22$; as well as to
the optimal fluctuations derived in the persistent limit.
This indicates that qualitative agreement reaches far into the non-persistent
(recall that in the~\DNSA~case, $St=1$ and $Ku=10.3$) regime and to
fully-resolved, high-Reynolds number turbulence.
Comparison of Figs.~\ref{fig:tracking}(a) to (b) and
Figs.~\ref{fig:tracking}(d) to (e), respectively, verify that
the conditions in terms of the fluctuation of $\A$ required to initiate a
caustic relax with increasing Stokes number.
In other words, heavier particles are driven into sling events more easily,
consistent with our observation of the regularizing effect of the $1/\tau_p$
term on the right-hand sides of \cref{eqn:1d,eqn:1dz}.
The most probable fluctuations, however, remain similar in shape.

Ahead of and after the sling event, particle trajectories are biased by
preferential sampling.
This is indicated by the fact that while $Q$ deviates from zero significantly
(relative to the peak of the fluctuation) outside the fluctuation, elevated $R$
is only seen in a narrow timespan before and shortly after the sling event.
In this regard, the notion of preferential sampling as a preconditioning
mechanism for caustics is convincing: The centrifuge causes particles to
preferentially sample $Q<0$ regions.
If, by chance, these regions coincide with $R>0$, particles are driven into a
sling event by propagating along the positive branch of the Vieillefosse line
(\cref{fig:QR}).
This is manifested by a fluctuation to positive $R$ and negative $Q$.
The $Q$-preconditionning is less intense at $St=0.2$, which makes caustics
rarer and requiring more extreme fluctuations of the FVG.

As mentioned earlier, our method cannot predict the \emph{exact} point in time
at which the blow up of $\tr(\Z)$ occurs.
The vertical lines in \cref{fig:tracking} merely depict the point in time at
which $D$ is measured compared to the threshold $D_c$.
Our method's uncertainty in matching the point in time at which the caustic
occurs is reflected by the artifacts of particles undergoing secondary
fluctuations visible in \cref{fig:tracking}(d).
These indicate that the sling event takes place before or after we record the
relative velocities.
The vast majority of particles, however, are selected at a reasonable time.
The variance is lower in the unfiltered cases, as the fluid timescales are much
shorter, whereby particles peak in relative velocity during a very brief period
of time.

Conspicuously, the sling particles of the \DNSB~case (\cref{fig:tracking}(c))
encounter extreme gradients from early on.
Almost all of them are initially located in regions of negative $Q$, with the
deviation from $R=0$ lagging slightly but still increasing early in the
process.
In other words, this leads one to suspect that for a particle with $St=5$, the only prerequisite for
participating in a sling event is to be initialized in a region of large
negative $Q$ at the start of the simulation and then being ejected into a
$R\gg0$ region shortly thereafter.
In truth, however, a history of FVGs that propagates along the Vieillefosse
line is required for a caustic to occur, irrespective of the Stokes number, as
shown in~\cref{fig:QR}.
At $St=5$, sling events take place on shorter timescales, yet are caused by
turbulent flow topologies in a manner very similar to the lower $St$ cases.
We do, however, expect that sling events can only occur below a certain Stokes
number, as particles become too heavy to respond to the flow at some point and
instead move ballistically.

In \cref{fig:QR}, it is observed that the gradients do not scatter on, but
slightly below the positive branch of the Vieillefosse tail.
An explanation for this circumstance was provided by
Meibohm~\etal~\cite{meibohmCausticFormationNonGaussian2024}.
In turbulence, strain and vorticity of the smallest scales is coupled in a way
such that the vorticity reduces the dissipation production, shifting $R$ to the
left.
Note that qualitatively similar joint probability densities were obtained in
the filtered cases, which are not shown in \cref{fig:QR} as they do not provide
any further insight.

\section{Conclusions}\label{sec:conclusion}
In this work, we have obtained results from extensive direct numerical
simulations of high-Reynolds number turbulence laden with large ensembles of
one-way coupled inertial point-particles.
The simulations were performed before and after applying a sharp spectral
filter to the flow fields.
The particles initially cluster where the local instantaneous strain dominates
the vorticity, imposing granularity on the spatial distribution that is
retained later on.
The filter is shown to not only modify this granularity, but to also have
strong effects on clustering by altering the relative importance of the
mechanisms at play.
While in the unfiltered cases, Maxey's centrifuge is found to be qualitatively
accurate, the filtered fields exhibit small-scale heterogeneity that can only
be explained by the path-history effect.
An effective Stokes number is used to compare unfiltered to filtered cases, and
it is shown that scale-similarity holds at similar (effective) Stokes numbers;
qualitatively in terms of the visual observations of the particle fields, and
quantitatively in terms of the fractal correlation dimension.
We then identified caustics by thresholding the relative velocities of close-by
particle pairs and provided one-, two-, and three-dimensional visualizations of
sling events, which occur as a result of rare, intense events of the turbulent
flow field.
Sling events were shown to take place in thin sheets, under compressive strain
and biased by preferential sampling of the flow topology of the fourth quadrant
in the $QR$-plane.
A characteristic optimal fluctuation in terms of $Q$ and $R$ was observed to
drive particles into sling events, suggesting that the recent results from
Meibohm~\etal~\cite{meibohmCausticsTurbulentAerosols2023,meibohmCausticFormationNonGaussian2024}
may extend to high-Reynolds turbulence beyond the persistent regime.
The separation of mechanisms indicates that preferential sampling is most
intense around $St=1$, whereas clustering by caustics peaks somewhere above
$St>1$.
These findings are, however, inconclusive and shall be investigated in more
detail in future work.

Regarding Euler-Lagrange large eddy simulations of particle-laden turbulence,
our results indicate that without models accounting for the sub-grid scales
seen by the particles, significant errors are incurred.
The absence of small-scale turbulence strongly alters the spatial distributions
of the particles and the rate at which caustics occur.
Clearly, sub-grid caustics cannot be resolved in LES.
In future work, we will exploit the scale-similarity of the particle fields in
the development of sub-grid scale models for LES: With filter-effective and
targeted Stokes numbers easily calculable, we envision LES modeling approaches
(similar to the thickened flame model for LES of turbulent premixed
combustion~\cite{colinThickenedFlameModel2000}), where the reconstruction of
the missing sub-grid structure is not attempted but accounted for through
corrections, e.g., to the effective collision, evaporation or condensation
rates, which depend on $St$ and $\Steff$.

\appendix*
\section{Backtracked trajectories} \label{sec:appendix}
\begin{figure*}
	\includegraphics[width=\textwidth]{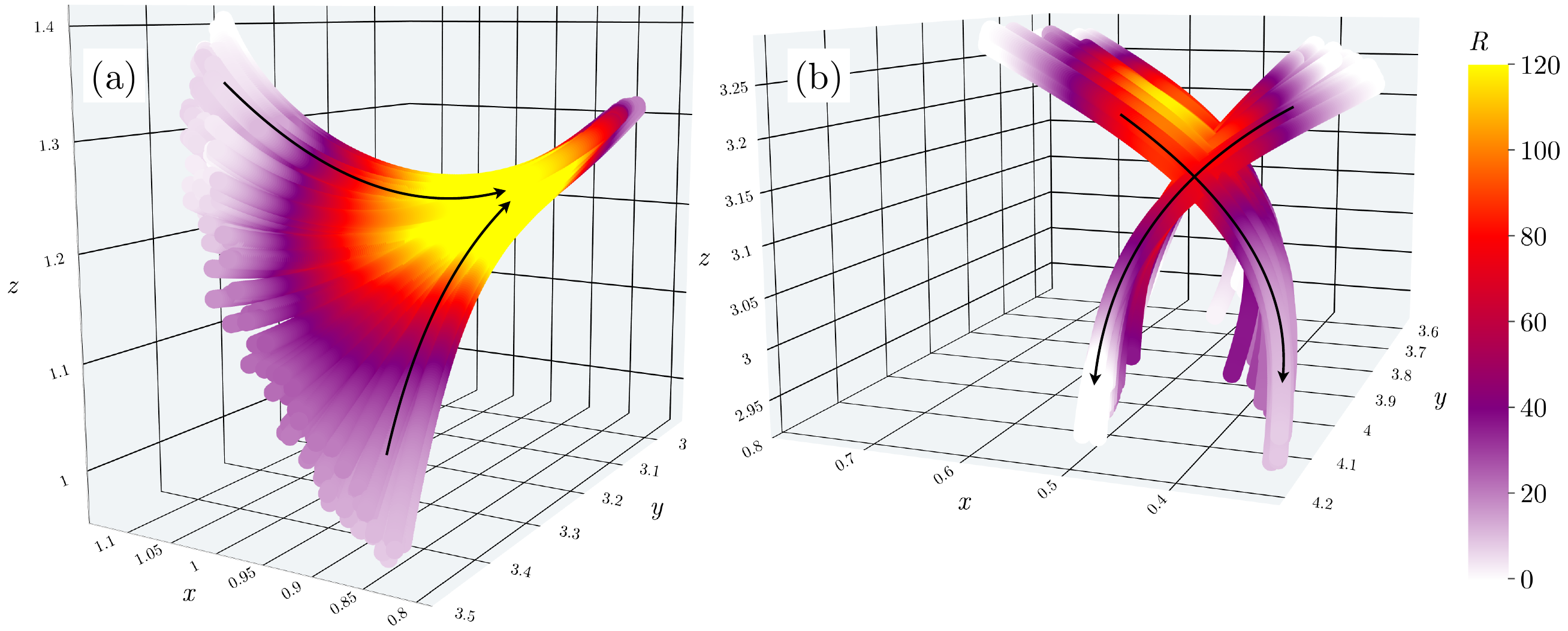}
	\caption{\label{fig:3dbacktrack}
		Trajectories of (a) \LESA~and~(b)~\LESB~sling particles.
		The ensemble was selected using the criterion $D>D_c$ at $t_c=0.6$ and
		$t_c=0.4$, respectively, and then tracked backward (and forward) in time to
		create the ranges $t\in[t_c-0.4,t_c]$ and $t\in[t_c-0.4,t_c+0.1]$.
		The colors depict the locally experienced value of $R$.
		The direction of the particle motion is denoted by arrows.
	}
\end{figure*}

Here, we provide additional 3D visualizations of sling events in order to
illustrate the back- and forward tracking that was used to produce
\cref{fig:3d,fig:tracking}.
Figures~\ref{fig:3dbacktrack}(a,b) show the trajectories of particles that engage in sling events
at $t_c=0.6$ (\LESA) and $t_c=0.4$ (\LESB), respectively.
The trajectories were obtained by tracking an ensemble of sling particles
backward and forward in time from $t=t_c$.
The color shows the fluctuation of local $R$ that is required for the sling
event.
Finally, we observe that at small Stokes number, the particles retain the form
of a thin sheet after the sling event has taken place, whereas a pocket forms
for $\Steff=0.98$, due to higher relative velocities.

\begin{acknowledgments}
	This work was funded by the Swiss National Science Foundation under grant
	200021\_204621.
	Numerical simulations were carried out using the resources provided by the
	\emph{Euler} cluster of ETH Z\"urich~\cite{ethzurichEuler}.
	We thank Heinrich Heinzer for his support.
\end{acknowledgments}

\bibliography{ref}

\end{document}